\documentclass{llncs}
\usepackage{llncsdoc}
\usepackage{booktabs}
\usepackage{url}
\usepackage{amssymb}
\usepackage{amsmath}
\usepackage{wrapfig}
\usepackage{enumitem}
\usepackage{mdwlist}
\usepackage[caption=false]{subfig}
\usepackage{microtype}
\usepackage{color}
\usepackage{capt-of}
\usepackage{algorithm}
\usepackage{tikz}
\usepackage{rotating}
\usepackage{makecell}
\usetikzlibrary{arrows,automata,backgrounds,decorations,fit,petri,positioning,petri,shapes,calc}
\tikzset{small/.style={level distance=10pt,sibling distance=-2pt,outer ysep=-2pt, label distance=-8pt}}
\tikzset{smaller/.style={level distance=20pt}}
\tikzset{smallish/.style={level distance=20pt,sibling distance=-2pt,outer ysep=-2pt, label distance=-8pt}}
\tikzset{node distance=5mm} 
\tikzset{place/.append style={circle,draw=black,thick,inner sep=0pt,minimum size=4mm,label position=below}} 
\tikzset{transition/.append style={rectangle,draw=black,thick,inner sep=0pt,minimum size=6.8mm}}
\tikzset{tau/.style={transition,fill=black}}  

\urldef{\mailsa}\path|{n.tax,n.sidorova}@tue.nl, {e.alasgarov}@student.tue.nl|
\urldef{\mailsb}\path|{niek.tax,reinder.haakma}@philips.com|
\urldef{\mailsc}\path|{e.alasgarov}@student.tue.nl|

\newcommand{\Lan}                 {\mathfrak{L}}
\newcommand{\N}                 {\mathbb{N}}
\newcommand{\E}                 {\mathcal{E}}
\newcommand{\pre}[1]{\bullet #1}

\begin{document}
\title{On Generation of Time-based Label Refinements}

\author{Niek Tax\inst{1,2} \and Emin Alasgarov\inst{1,2} \and Natalia Sidorova\inst{1} \and Reinder Haakma\inst{2}}

\institute{Eindhoven University of Technology, Department of Mathematics and Computer Science, P.O. Box 513, 5600MB Eindhoven, The Netherlands\\
\mailsa
\and Philips Research, Prof. Holstlaan 4, 5665 AA Eindhoven, The Netherlands\\
\mailsb
}

\maketitle
\begin{abstract}
Process mining is a research field focused on the analysis of event data with the aim of extracting insights in processes. Applying process mining techniques on data from smart home environments has the potential to provide valuable insights in (un)healthy habits and to contribute to ambient assisted living solutions. Finding the right event labels to enable application of process mining techniques is however far from trivial, as simply using the triggering sensor as the label for sensor events results in uninformative models that allow for too much behavior (overgeneralizing). Refinements of sensor level event labels suggested by domain experts have shown to enable discovery of more precise and insightful process models. However, there exist no automated approach to generate refinements of event labels in the context of process mining. In this paper we propose a framework for automated generation of label refinements based on the time attribute of events. We show on a case study with real life smart home event data that behaviorally more specific, and therefore more insightful, process models can be found by using automatically generated refined labels in process discovery.
\end{abstract}

\keywords{Label Refinements, Process Discovery, Unsupervised Learning}

\section{Introduction}
Process mining is a fast growing discipline that combines knowledge and techniques from data mining, process modeling, and process model analysis~\cite{Aalst2016}. Process mining techniques concern the analysis of events that are logged during process execution, where event records contain information on what was done, by whom, for whom, where, when, etc. Events are grouped into cases (process instances), e.g. per patient for a hospital log, or per insurance claim for an insurance company. \emph{Process discovery} plays an important role in process mining, focusing on extracting interpretable models of processes from event logs. One of the attributes of the events is usually used as its label and its values become transition/activity labels in the process models generated by process discovery algorithms.\looseness=-1

The scope of process mining have broadened in recent years from business process management to other application domains, one of them being analysis of events of human behavior with data originating from sensors in smart home environments \cite{Sztyler2015,Tax2016a,Tax2016b}. Table \ref{tab:example_log} shows an example of such an event log. Events in the event log are generated by e.g. motion sensors placed in the home, power sensors placed on appliances, open/close sensors placed on closets and cabinets, etc. Particularly challenging in applying process mining in this application domain is the extraction of meaningful event labels that allow for discovery of insightful process models. Simply using the sensor that generates an event (the \emph{sensor} column in Table \ref{tab:example_log}) as event label is shown to produce non-informative process models that overgeneralize the event log and allow for too much behavior \cite{Tax2016a}. Abstracting sensor-level events into events at the level of human activity (e.g. \emph{eating}, \emph{sleeping, etc.}) using techniques closely related to techniques used in the activity recognition field helps to discover more behaviorally more constrained and insightful process models \cite{Tax2016b}, but applicability of this approach relies on the availability of a reliable diary of human behavior at the activity level, which is often just impossible to obtain.\looseness=-1

\begin{table}[t]
\centering\scriptsize
\caption{An example of an event log from a smart home environment.}
\scalebox{0.7}{
	\begin{tabular}{ccccc}
		\toprule
		Id & Timestamp & Address & Sensor & Sensor value\\
		\midrule
		1 & \textcolor{gray}{03/11/2015} 04:59:54 & \textcolor{gray}{Mountain Rd. 7} & Motion sensor - Bedroom  &  1\\
		2 & \textcolor{gray}{03/11/2015} 06:04:36 & \textcolor{gray}{Mountain Rd. 7} & Motion sensor - Bedroom & 1\\
		3 & \textcolor{gray}{03/11/2015} 08:45:12 & \textcolor{gray}{Mountain Rd. 7} & Motion sensor - Living room &  1\\
		4 & \textcolor{gray}{03/11/2015} 09:10:10 & \textcolor{gray}{Mountain Rd. 7} & Motion sensor - Kitchen & 1\\
		5 & \textcolor{gray}{03/11/2015} 09:12:01 & \textcolor{gray}{Mountain Rd. 7} & Power sensor - Water cooker & 1200\\
		6 & \textcolor{gray}{03/11/2015} 09:15:45 & \textcolor{gray}{Mountain Rd. 7} & Power sensor - Water cooker & 0\\
		\dots & \textcolor{gray}{03/11/2015} \dots & \textcolor{gray}{Mountain Rd. 7} & \dots & \dots \\
		\midrule
		7 & \textcolor{gray}{03/12/2015} 01:01:23 & \textcolor{gray}{Mountain Rd. 7} & Motion sensor - Bedroom &  1\\
		8 & \textcolor{gray}{03/12/2015} 03:13:14 & \textcolor{gray}{Mountain Rd. 7} & Motion sensor - Bedroom &  1\\
		9 & \textcolor{gray}{03/12/2015} 07:24:57 & \textcolor{gray}{Mountain Rd. 7} & Motion sensor - Bedroom &  1\\
		10 & \textcolor{gray}{03/12/2015} 08:34:02 & \textcolor{gray}{Mountain Rd. 7} & Motion sensor - Bedroom &  1\\
		11 & \textcolor{gray}{03/12/2015} 09:12:00 & \textcolor{gray}{Mountain Rd. 7} & Motion sensor - Living room & 1\\
		\dots & \textcolor{gray}{03/12/2015} \dots & \textcolor{gray}{Mountain Rd. 7} & \dots & \dots \\
		\midrule
		12 & \textcolor{gray}{03/14/2015} 03:41:46 & \textcolor{gray}{Mountain Rd. 7} & Motion sensor - Bedroom & 1\\
		13 & \textcolor{gray}{03/14/2015} 05:00:17 & \textcolor{gray}{Mountain Rd. 7} & Motion sensor - Bedroom & 1\\
		14 & \textcolor{gray}{03/14/2015} 08:52:32 & \textcolor{gray}{Mountain Rd. 7} & Motion sensor - Bedroom & 1\\
		15 & \textcolor{gray}{03/14/2015} 09:30:54 & \textcolor{gray}{Mountain Rd. 7} & Motion sensor - Living room & 1\\
		16 & \textcolor{gray}{03/14/2015} 09:35:25 & \textcolor{gray}{Mountain Rd. 7} & Power sensor - TV & 160 \\
		17 & \textcolor{gray}{03/14/2015} 10:27:37 & \textcolor{gray}{Mountain Rd. 7} & Power sensor - TV & 0 \\
		\dots & \textcolor{gray}{03/14/2015} \dots & \textcolor{gray}{Mountain Rd. 7} & \dots & \dots \\
		\midrule
		\dots&\dots&\dots&\dots&\dots\\
		\bottomrule
	\end{tabular}}
	\label{tab:example_log}
	\vspace{-0.4cm}
\end{table}

In our earlier work \cite{Tax2016a} we showed that better process models can be discovered by taking the name of the sensor that generated the event as a starting point for the event label and then refining these labels using information on the time within the day at which the event occurred. The refinements used in \cite{Tax2016a} were based on domain knowledge, and not identified automatically from the data. In this paper, we aim at automatic generation of semantically interpretable label refinements that can be explained to the user, by basing label refinements on data attributes of events. We explore methods to bring parts of the timestamp information to the event label in an intelligent and fully automated way, with the end goal of discovering behaviorally more precise and therefore more insightful process models.\looseness=-1

We start by introducing basic concepts and notations used in this paper in Section \ref{sec:preliminaries}. In Section \ref{sec:approach}, we introduce a framework for the generation of event labels refinements based on the time attribute. In Section \ref{sec:case_study}, we apply this framework on a real life smart home data set and show the effect of the refined event labels on process discovery.  We continue by describing related work in Section \ref{sec:related_work} and conclude in Section \ref{sec:conclusion}.\looseness=-1

\section{Preliminaries}
\label{sec:preliminaries}
In this section we introduce the notions related to event logs and relabeling functions for traces and then define the notions of refinements and abstractions. We also introduce the Petri net process model notation.

We use the usual sequence definition, and denote a sequence by listing its elements, e.g. we write $\langle a_1,a_2,\dots,a_{n} \rangle$ for a (finite) sequence $s:\{1,\dots,n\}\to A$ of elements from some alphabet $A$, where $s(i)=a_i$ for any $i \in \{1,\dots,n\}$; $|s|$ denotes the length of sequence $s$; $s_1 s_2$ denotes the concatenation of sequences $s_1$ and $s_2$. A \emph{language} $\Lan$ over an alphabet $A$ is a set of sequences over $A$. $\Lan^p$ is the prefix closure of a language $\Lan$ (with $\Lan\subseteq\Lan^p$).

An event is the most elementary element of an event log. Let $\mathcal{I}$ be a set of event identifiers, and $\mathcal{A}_1 \times \dots \times \mathcal{A}_{n}$ be an attribute domain consisting of $n$ attributes (e.g. timestamp, resource, activity name, cost, etc.). An event is a tuple $e=(i,a_1,\dots,a_{n})$, with $i\in\mathcal{I}$ and $(a_1, \dots, a_{n})\in \mathcal{A}_1 \times \dots \times \mathcal{A}_{n}$. The \emph{event label} of an event is the attribute set $(a_1\dots,a_n)$; $e_i$, and $e_a$ respectively denote the identifier and label of event $e$. The timestamp attribute of an event is denoted by $a_t$. $\mathcal{E}=\mathcal{I}\times \mathcal{A}_1 \times \dots \times \mathcal{A}_{n}$ is a universe of events over $\mathcal{A}_1, \dots, \mathcal{A}_{n}$.
The rows of Table \ref{tab:example_log} are events from an event universe over the event attributes \emph{timestamp}, \emph{sensor}, \emph{address}, and \emph{sensor value}.

Events are often considered in the context of other events. We call $E\subseteq\mathcal{E}$ an \emph{event set} if $E$ does not contain any events with the same event identifier. The events in Table \ref{tab:example_log} together form an event set. A \emph{trace} $\sigma$ is a finite sequence formed by the events from an event set $E\subseteq{\mathcal{E}}$ that respects the time ordering of events, i.e. for all $k,m\in\N$, $1\leq k < m \leq |E|$, we have: $\sigma(k)_t\leq \sigma(m)_t$. We define the \emph{universe of traces} over event universe $\mathcal{E}$, denoted $\Sigma(\mathcal{E})$, as the set of all possible traces over $\mathcal{E}$. We omit $\mathcal{E}$ in $\Sigma(\mathcal{E})$ and use the shorter notation $\Sigma$ when the event universe is clear from the context.\looseness=-1 

Often it is useful to partition an event set into smaller sets in which events belong together according to some criterion. We might for example be interested in discovering the typical behavior within a household over the course of a day. In order to do so, we can e.g.~group together events with the same \emph{address} and the same day-part of the \emph{timestamp}, as indicated by the horizontal lines in Table \ref{tab:example_log}. For each of these event sets, we can construct a trace; time stamps define the ordering of events within the trace. For events of a trace having the same time stamps, an arbitrary ordering can be chosen within a trace.

An \emph{event partitioning function} is a function $ep: \mathcal{E} \to T_{id}$ that defines the partitioning of an arbitrary set of events $E\subseteq\mathcal{E}$ from a given event universe $\mathcal{E}$ into event sets $E_1,\ldots,E_j,\ldots$ where each $E_j$ is the maximal subset of $E$ such that for any $e_1, e_2\in E_j$, $ep(e_1)= ep(e_2)$; the value of $ep$ shared by all the elements of $E_j$ defines the value of the \emph{trace attribute} $T_{id}$. Note that multidimensional trace attributes are also possible, i.e. a combination of the name of the person performing the event activity and the date of the event, so that every trace contains activities of one person during one day. The event sets obtained by applying an event partitioning can be transformed into traces (respecting the time ordering of events).\looseness=-1


An event log $L$ is a finite set of traces $L \subseteq \Sigma(\mathcal{E})$. $A_L\subseteq\mathcal{A}_1 \times \dots \times \mathcal{A}_{n}$ denotes the \emph{alphabet of event labels} that occur in log $L$. The traces of a log are often transformed before doing further analysis: very detailed but not necessarily informative event descriptions are transformed into some \emph{informative} and \emph{repeatable} labels. For the labels of the log in Table~\ref{tab:example_log}, the sensor values could be abstracted to \textit{on}, and \textit{off} or labels can be redefined to a subset of the event attributes, e.g. leaving the sensor values out completely. Next to that, if the event partitioning function maps each event from Table~\ref{tab:example_log} to its address and the day-part of the timestamp, these attributes (indicated in gray) become the trace attribute and can safely be removed from individual events.

After this relabeling step, some traces of the log can become identically labeled (the event id's would still be different). The information about the number of occurrences of a sequence of labels in an event log is highly relevant for process mining, since it allows differentiating between the mainstream behavior of a process (frequently occurring behavioral patterns) and exceptional behavior.

Let $\E, \E'$ be an event universe. A function $l: \E \to \E'$ is an \emph{event relabeling function}. A relabeling function can be used to obtain more useful event labels than the full set of event attribute values. We lift $l$ to event logs. Let $\E,\E_1,\E_2$ be event universes with $\E,\E_1,\E_2$ being pairwise different. Let $l_1: \E \to \E_1$ and $l_2: \E \to \E_2$ be event relabeling functions. Relabeling function $l_1$ is a \emph{refinement} of relabeling function $l_2$, denoted by $l_1\preceq l_2$, iff $\forall_{e_1,e_2\in \E}:l_1(e_1)=l_1(e_2)\implies l_2(e_1)=l_2(e_2)$; $l_2$ is then called an \emph{abstraction} of $l_1$.

The goal of process discovery is to discover a process model that represents the behavior seen in an event log. A frequently used process modeling notation in the process mining field is the Petri net~\cite{Reisig1998}. Petri nets are directed bipartite graphs consisting of transitions and places, connected by arcs. Transitions represent activities, while places represent the enabling conditions of transitions. Labels are assigned to transitions to indicate the type of activity that they model. A special label $\tau$ is used to represent invisible transitions, which are only used for routing purposes and not recorded in the execution log.

\begin{figure}[t]
	\centering
	\scalebox{0.67}{
		\begin{tikzpicture}
		[node distance=1.11cm,
		on grid,>=stealth',
		bend angle=20,
		auto,
		every place/.style= {minimum size=4mm},
		every transition/.style = {minimum size = 8.5mm}
		]
		\node [place, tokens = 1] (p1) [label=below:$p_1$]{};
		\node [transition] (b) [label=below:$t_1$, right = of p1] {$A$}
		edge [pre] node[auto] {} (p1);
		\node [place] (p2) [label=below:$p_2$, right = of b]{}
		edge [pre] node[auto] {} (b);
		\node [transition] (c) [label=below:$t_3$, below right = of p2] {$C$}
		edge [pre] node[auto] {} (p2);
		\node [transition] (c2) [label=below:$t_2$, above right = of p2] {$B$}
		edge [pre] node[auto] {} (p2);
		\node [place] (p3) [label=below:$p_3$, above right = of c]{}
		edge [pre] node[auto] {} (c)
		edge [pre] node[auto] {} (c2);
		\node [transition] (d) [label=below:$t_4$, right = of p3] {$D$}
		edge [pre] node[auto] {} (p3);
		\node [place] (p4) [label=above:$p_4$, below right = of d]{}
		edge [pre] node[auto] {} (d);
		\node [transition] (e) [label=above:$t_5$, right = of p4] {$F$}
		edge [pre] node[auto] {} (p4);
		\node [place] (p6) [label=above:$p_6$, right = of e]{}
		edge [pre] node[auto] {} (e);
		\node [place] (p5) [label=above:$p_5$, above right = of d]{}
		edge [pre] node[auto] {} (d);
		\node [transition] (f) [label=above:$t_6$, right = of p5] {$E$}
		edge [pre] node[auto] {} (p5);
		\node [place] (p7) [label=above:$p_7$, right = of f]{}
		edge [pre] node[auto] {} (f);
		\node [transition] (g) [label=below:$t_7$, minimum width=3mm, fill=lightgray, below right = of p7] {}
		edge [pre] node[auto] {} (p7)
		edge [pre] node[auto] {} (p6);
		\node [place] (p8) [label=below:$p_8$, right = of g]{}
		edge [pre] node[auto] {} (g);
		
		\end{tikzpicture}}
	\caption{An example Petri net.}
	\label{fig:example_petri_net}
	\vspace{-0.25cm}
\end{figure}
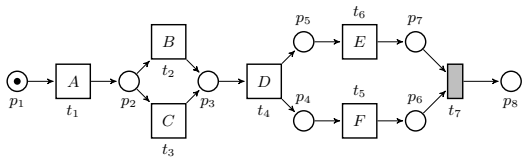

A \emph{labeled Petri net} $N=\langle P,T,F,A_M,\ell\rangle$ is a tuple where $P$ is a finite set of places, $T$ is a finite set of transitions such that $P \cap T = \emptyset$,  $F \subseteq (P \times T) \cup (T \times P)$ is a set of directed arcs, called the flow relation, $A_M$ is an alphabet of labels representing activities, with $\tau \notin A_M$ being a label representing  invisible events, and $\ell:T\rightarrow A_M\cup \{\tau\}$ is a labeling function that assigns a label to each transition. For a node $n \in (P \cup T)$ we use $\bullet n$ and $n \bullet$ to denote the set of input and output nodes of $n$, defined as $\bullet n =\{n|(n',n)\in F\}$ and $n \bullet =\{n|(n,n')\in F\}$. An example of a Petri net can be seen in Figure \ref{fig:example_petri_net}, where circles represent places and squares represent transitions. Gray transitions with smaller width represent $\tau$ transitions.\looseness=-1

A state of a Petri net is defined by its \emph{marking} $M \in \mathbb{N}^{P}$ being a multiset of places. A marking is graphically denoted by putting $M(p)$ tokens on each place $p\in P$. A pair $(N,M)$ is called a marked Petri net. State changes occur through transition firings. A transition $t$ is enabled (can fire) in a given marking $M$ if each input place $p\in \pre{t}$ contains at least one token. Once a transition fires, one token is removed from each input place of $t$  and one token is added to each output place of $t$, leading to a new marking. An \emph{accepting Petri net} is a 3-tuple $(N,M_i,M_f)$ with $N$ a labeled Petri net, $M_i$ an initial marking, and $M_f$ a set of final markings.
Many process modeling notations, including accepting Petri nets, have formal executional semantics and a model defines a \emph{language of accepting traces} $\Lan$. For the Petri net in Figure \ref{fig:example_petri_net}, the language of accepting traces is $\{\langle A,B,D,E,F\rangle,\langle A,B,D,F,E\rangle,\langle A,C,D,E,F\rangle,\langle A,C,D,F,E\rangle\}$.

\section{A Framework for Time-based Label Refinements}
\label{sec:approach}
To generate potential label refinements for every label based on time we take a clustering based approach by identifying dense areas in time space for each label. The time part of the timestamps consists of values between $\textit{00:00:00}$ and $\textit{23:59:59}$, equivalent to the timestamp attribute from Table \ref{tab:example_log} with the day-part of the timestamp removed. This timestamp can be transformed into a real number hourfloat representation in interval $[0,24)$. We chose to apply soft clustering (also referred to as fuzzy clustering), which has the benefit of assigning to each data point a likelihood of belonging to each cluster. A well-known approach to soft clustering is based on the combination of the Expectation-Maximization (EM) algorithm with mixture models, which are probability distributions consisting of multiple components of the same probability distribution. Each component in the mixture represents one cluster and the probability of a data point belonging to that cluster is the probability that this cluster generated that data point. The EM algorithm is used to obtain a maximum likelihood estimate of the mixture model parameters, i.e. the parameters of the probability distributions in the mixture.\looseness=-1

\begin{figure}[t]
	\centering
	\includegraphics[width=0.675\linewidth]{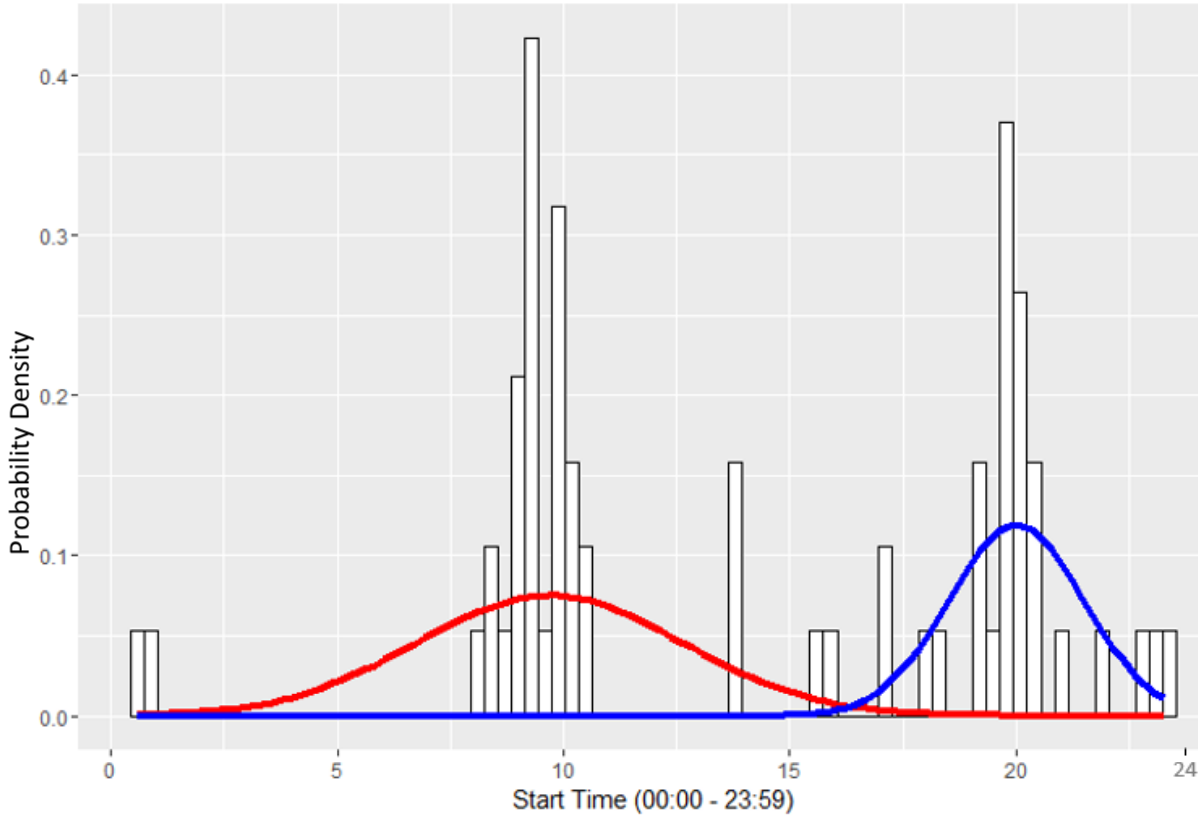}
	\caption{The histogram representation and a Gaussian Mixture Model fitted to timestamps values of the plates cupboard sensor in the van Kasteren data set \cite{Kasteren2008}.\looseness=-1}
	\label{fig:problem_data_set}
	\vspace{-0.2cm}
\end{figure}

 A well-known example of a mixture model is the Gaussian Mixture Model (GMM), where the components in the mixture distributions are normal distributions. The data space of time is, however, non-euclidean: it has a circular nature, e.g. $23.99$ is closer to $0$ than to $23$. This circular nature of the data space introduces problems for GMMs, as shown in Figure \ref{fig:problem_data_set}. The GMM fitted to the timestamps of the sensor events consists of two components, one with the mean at $9.05$  and one with a mean at $20$. The histogram representation of the same data shows that some events happened just after midnight, which is actually closer on the clock to $20$ than to $9.05$. The GMM however is unaware of the circularity of the clock, which results  in the mixture model that seems inappropriate when visually comparing with the histogram. The field of circular statistics (also referred to as directional statistics), concerns analysis of such circular data spaces (cf. \cite{Mardia2009}).\looseness=-1

Here, we introduce a framework for generating refinements of event labels based on time attributes using techniques from the field of circular statistics. This framework consists of three stages:
\begin{description}
	\item[Data-model pre-fitting stage]{A known problem with many clustering techniques is that they return clusters even when the data should not be clustered. In this stage we assess how many clusters the events of a sensor type contain.}
	\item[Data-model fitting stage]{In this stage we cluster the events of a sensor type by timestamp using a mixture consisting of components that take into account the circularity of the data.}
	\item[Data-model post-fitting stage]{In this stage the quality of the label refinements is assessed from both a cluster quality perspective and a process model (event ordering statistics) perspective.}
\end{description}


\subsection{Data-model pre-fitting stage}
We now describe a test for uniformity, a test for unimodality, and a method to select the number of clusters in the data.
\subsubsection{Uniformity Check - Rao's Spacing Test}
Rao's spacing test~\cite{Rao1976} tests the uniformity of the timestamps of the events from a sensor around the circular clock. This test is based on the idea that uniform circular data is distributed evenly around the circle, and $n$ observations are separated from each other $\frac{360}{n}$ degrees. The null hypothesis is that the data is uniform around the circle.

Given $n$ successive observations $f_1,\ldots,f_n$, either clockwise or counterclockwise, the test statistics $U$ for Rao's Spacing Test is defined as $U = \frac{1}{2}\sum_{i = 1}^{n}\mid T_i - \lambda \mid$, where $\lambda = \frac{360^\circ}{n}$, $T_i = f_{i + 1} - f_{i}$ for $1 \le i \le n - 1$ and $T_n = (360^\circ - f_n) + f_1$.

\subsubsection{Unimodality Check - Hartigan's Dip Test}
Hartigan's dip tests~\cite{Hartigan1985} the null hypothesis that the data follows a unimodal distribution on a circle. When the null hypothesis can be rejected, we know that the distribution of the data is at least bimodal. Hartigan's dip test measures the maximum difference between the the empirical distribution function and the unimodal distribution function that minimizes that maximum difference.

\subsubsection{Number of Component Selection - Bayesian Information Criterion}
The Bayesian Information Criterion (BIC)~\cite{Schwarz1978} introduces a penalty for the number of model parameters to the evaluation of a mixture model. Adding a component to a mixture model increases the number of parameters of the mixture with the number of parameters of the distribution of the added component. The likelihood of the data given the model can only increase by adding extra components, adding the BIC penalty results in a trade-off between number of components and the likelihood of the data given the mixture model. BIC is formally defined as $\textit{BIC} = -2 * ln\hat{L} + k * ln(n)$, where $\hat{L}$ is a maximized value for the data likelihood, $n$ is the sample size, and $k$ is the number of parameters to be estimated. A lower BIC value indicates a better model. We start with $1$ component, and iteratively increase from $k$ to $k+1$ components as long as the decrease in BIC is larger than 10, which is the threshold for decisive evidence of high BIC \cite{Kass1995}.

\subsection{Data-model fitting stage}
We cluster events generated by one sensor using a mixture model consisting of components of the von Mises distribution, which is a circular version of the normal distribution. This technique is based on the approach of Banerjee et al. \cite{Banerjee2005}, who introduce a clustering method based on a mixture of von Mises-Fisher distribution components, which is a generalization of the $2$-dimensional von Mises distribution to $n$-dimensional spheres. A probability density function for a von Mises distribution with mean direction $\mu$ and concentration parameter $\kappa$ is defined as $\textit{pdf}(\theta \mid \mu, \kappa) = \frac{1}{2\pi I_0(\kappa)}e^{\kappa\cos(\theta - \mu)}$, where mean $\mu$ and data point $\theta$ are expressed in radians on the circle, such that $0 \le \theta \le 2\pi, ~0 \le \mu \le 2\pi, ~\kappa \ge 0$.
\textit{$I_0$} represents the modified Bessel function of order 0, defined as $I_0(k) = \frac{1}{2\pi}\int_0^{2\pi} e^{\kappa\cos(\theta)}d\theta$. As $\kappa$ approaches 0, the distribution becomes uniform around the circle. As $\kappa$ increases, the distribution becomes relatively concentrated around the mean $\mu$ and the von Mises distribution starts to approximate a normal distribution. We fit a mixture model of von Mises components using the package movMF \cite{Hornik2014} provided in R.

\subsection{Data-model post-fitting stage}
After fitting a mixture of von Mises distributions to the sensor events, we perform a goodness-of-fit test to check whether the data could have been generated from this distribution. We describe the Watson $U^2$ statistic~\cite{Watson1962}, a goodness-of-fit assessment based on hypothesis testing.
The Watson $U^2$ statistic measures the discrepancy between the cumulative distribution function $F(\theta)$ and the empirical distribution function $F_n(\theta)$ of some sample $\theta$ drawn from some population and is defined as $U^2 = n\int_0^{2\pi} \Big[ F_n(\theta) - F(\theta) - \int_0^{2\pi} \big\{ F_n(\phi) - F(\phi) \big\} dF(\phi) \Big]^2 dF(\theta)$.

Furthermore we assess the quality of refining the event label into a new label for each cluster from a process perspective using the label refinement evaluation method described in \cite{Tax2016a}. This method tests whether the log statistics that are used in many process discovery algorithms become significantly more deterministic by applying the label refinement.

\section{Case Study}
\label{sec:case_study}

\begin{figure}[t]
	\centering
	\begin{minipage}[t]{.5\textwidth}
	\centering
	\includegraphics[width=\linewidth]{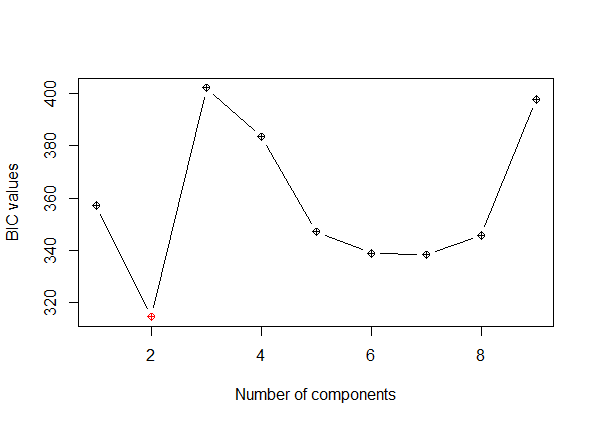}
	\caption{BIC values for different numbers of components in the mixture model.}
	\label{fig:num_component_selection_kasteren}
	\end{minipage}\hfill
	\begin{minipage}[t]{.4\textwidth}
		\centering
		\vspace{-3.9cm}
		\captionof{table}{Estimated parameters for a mixture of von Mises components for bedroom door sensor events.\looseness=-1}
		\label{tab:bedroom_components}
		\vspace{0.5cm}
		\begin{tabular}{c|ccc}
			\toprule
			Cluster & $\alpha$ & $\mu$ (radii) & $\kappa$ \\
			\midrule
			Cluster 1 & 0.76 & 2.05 & 3.85\\
			Cluster 2 & 0.24 & 5.94 & 1.56\\
			\bottomrule
		\end{tabular}
	\end{minipage}
	\vspace{-0.2cm}
\end{figure}

\begin{figure}[t]
	\centering
	\subfloat[]{
		\includegraphics[width=0.98\linewidth]{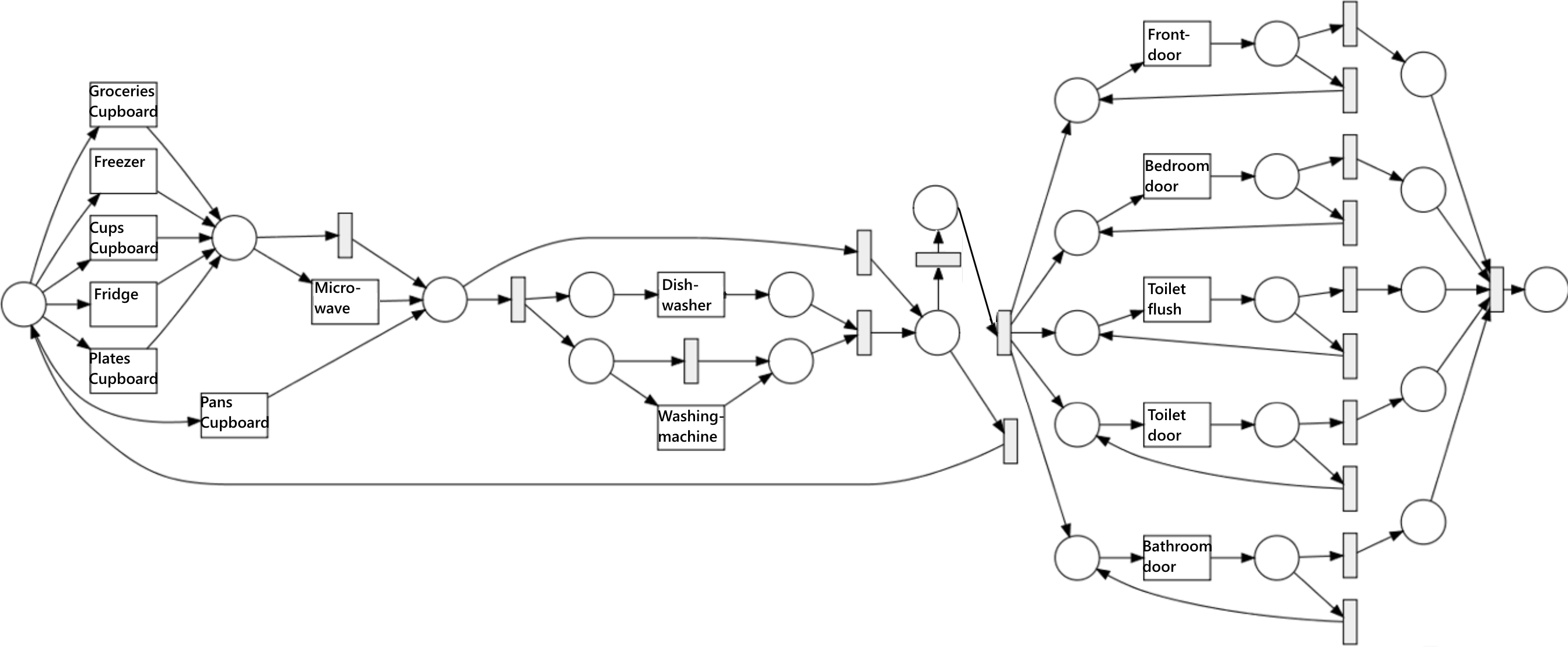}
		\label{fig:kasteren_original}
	}\\
	\subfloat[]{
		\hspace{-0.025\linewidth}
		\includegraphics[width=1.0125\linewidth]{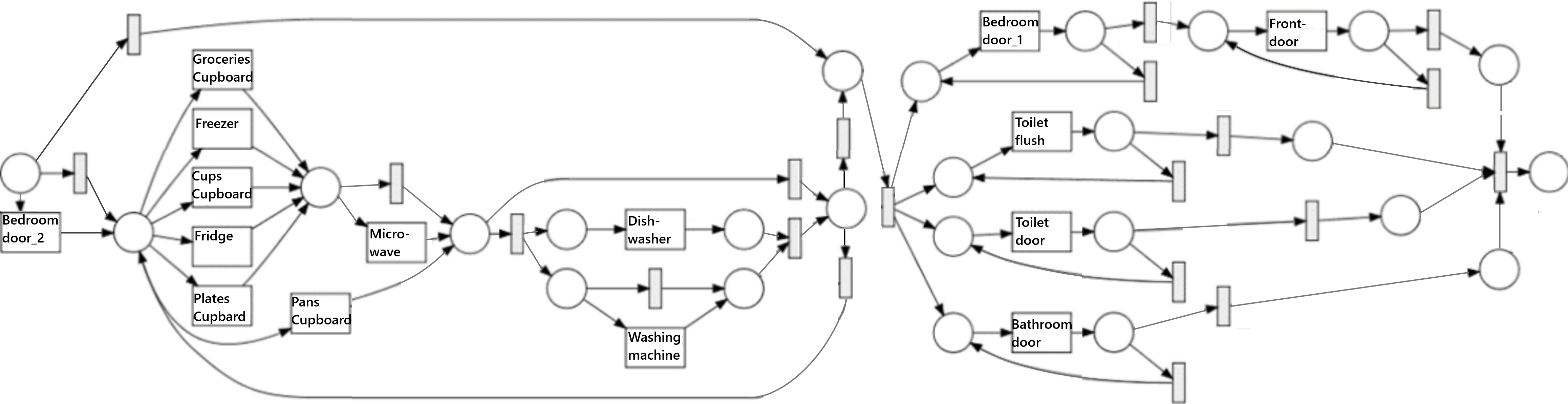}
		\label{fig:kasteren_refined}
	}
	\caption{Process models discovered on the van Kasteren data with sensor-level labels (a) and refined labels (b) with the Inductive Miner infrequent (20\% filtering) \cite{Leemans2013}.}
	\vspace{-0.2cm}
\end{figure}

\begin{figure}
	\centering
	\hspace{-0.025\linewidth}
	\includegraphics[width=1.0125\linewidth]{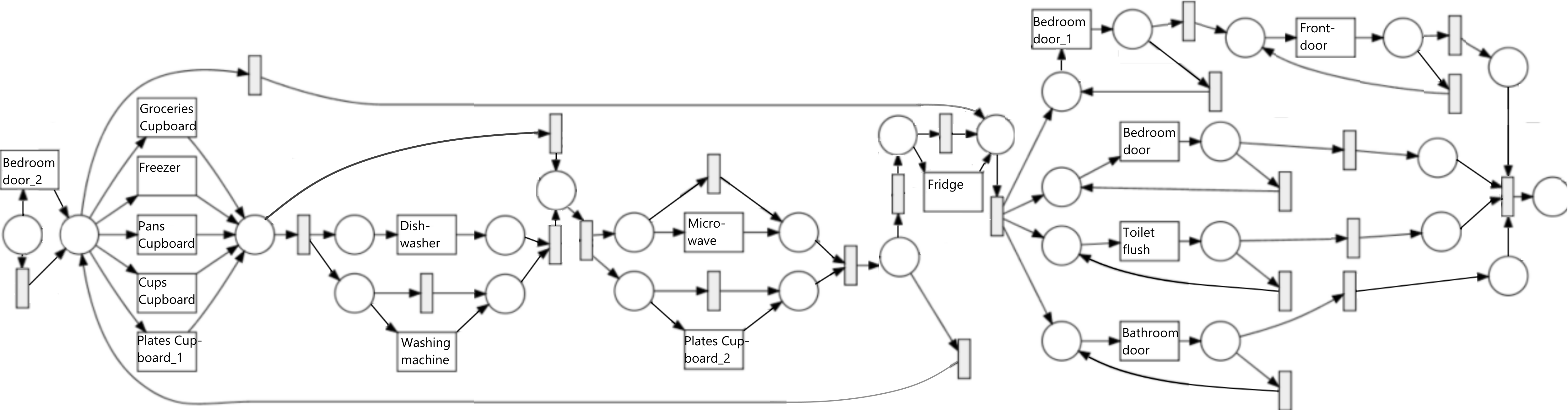}
	\caption{Inductive Miner infrequent (20\% filtering) \cite{Leemans2013} result after a second label refinement.}
	\label{fig:kasteren_refined2}
	\vspace{-0.2cm}
\end{figure}

We show the results of our time-based label refinements approach on the real life smart home data set described in van Kasteren et al. \cite{Kasteren2008}. The van Kasteren data set consists of 1285 events divided over fourteen different sensors. We segment in days from midnight to midnight to define cases. Figure \ref{fig:kasteren_original} shows the process model discovered on this event log with the Inductive Miner infrequent \cite{Leemans2013} with 20\% filtering, which discovers a process model that describes the most frequent 80\% of behavior in the log. Note that this process model overgeneralises allowing too much behaviour. At the beginning a (possibly repeated) choice is made between five transitions. At the end of the process, the model allows any sequence over the alphabet of  five activities, where each activity occurs at least once.

We illustrate our proof of concept by applying the framework to the \emph{bedroom door} sensor. Rao's spacing test results in a test statistic of $241.0$ with $152.5$ being the critical value for significance level $0.01$, indicating that we can reject the null hypothesis of a uniformly distributed set of \emph{bedroom door} timestamps. Hartigan's dip test results in a p-value of $3.95\times10^{-4}$, indicating that we can reject the null hypothesis that there is only one cluster in the \emph{bedroom door} data. Figure \ref{fig:num_component_selection_kasteren} shows the BIC values for different numbers of components in the model. The figure indicates that there are two clusters in the data, as this corresponds to the lowest BIC value. Table \ref{tab:bedroom_components} shows the mean and $\kappa$ parameters of the two clusters found by optimizing the von Mises mixture model with the EM algorithm. A value of $0=2\pi$ radii equals midnight. After applying the von Mises mixture model to the \emph{bedroom door} events and assigning each event to the maximum likelihood cluster we obtain a time range of [3.08-10.44] for cluster 1 and a time range of [17.06-0.88] for cluster 2. The Watson $U^2$ test results in a test statistic of $0.368$ and $0.392$ for cluster 1 and 2 respectively with a critical value of $0.141$ for a $0.01$ significance level, indicating that the data is likely to be generated by the two von Mises distributions found. The label refinement evaluation method \cite{Tax2016a} finds statistically significant differences between the events from the two \emph{bedroom door} clusters with regard to their control-flow relations with other activities in the log for 10 other activities using the significance level of $0.01$, indicating that the two clusters are different from a control-flow perspective. Figure \ref{fig:kasteren_refined} shows the process model discovered with the Inductive Miner infrequent with 20\% filtering after applying this label refinement to the van Kasteren event log. The process model still overgeneralizes in general, but the label refinement does help restricting the behavior, as it shows that the evening \emph{bedroom door} events are succeeded by one or more events of type \emph{groceries cupboard}, \emph{freezer}, \emph{cups cupboard}, \emph{fridge}, \emph{plates cupboard}, or \emph{pans cupboard}, while the morning \emph{bedroom door} events are followed by one or more \emph{frontdoor} events. It seems that this person generally goes to the bedroom in-between coming home from work and starting to cook. The loop of the \emph{frontdoor} events could be caused by the person leaving the house in the morning for work, resulting in no logged events until the person comes home again by opening the \emph{frontdoor}. Note that in Figure \ref{fig:kasteren_original} \emph{bedroom door} and \emph{frontdoor} events can occur an arbitrary number of times in any order. Figure \ref{fig:kasteren_original} furthermore does not allow for the \emph{bedroom door} to occur before the whole block of kitchen-located events at the beginning of the net. \looseness=-1

Label refinements can be applied iteratively. Figure \ref{fig:kasteren_refined2} shows the effect of a second label refinement step, where \emph{Plates cupboard} using the same methodology is refined into two labels, representing time ranges [7.98-14.02] and [16.05-0.92] respectively. This refinement shows the additional insight that the evening version of the \emph{Plates cupboard} occurs in directly before or after the microwave.

\section{Related Work}
\label{sec:related_work}
Refining event labels in the event log is closely related to the task of mining process models with duplicate activities, in which the resulting process model can contain multiple transitions/nodes with the same label. From the point of view of the behavior allowed by a process model, it makes no difference whether a process model is discovered on an event log with refined labels, or whether a process model is discovered with duplicate activities such that each transition/node of the duplicate activity precisely covers one versions of the refined label. The first process discovery algorithm capable of discovering duplicate tasks was proposed by Herbst and Karagiannis in 2004 \cite{Herbst2004}, after which many others have been proposed, including the Genetic Miner \cite{DeMedeiros2007}, the Evolutionary Tree Miner \cite{Buijs2012}, the $\alpha^*$-algorithm \cite{Li2007}, the $\alpha^\#$-algorithm \cite{Gu2008}, the EnhancedWFMiner \cite{Folino2009}, and a simulated annealing based algorithm \cite{Song2008}. An alternative approach has been proposed by V\'{a}zques-Barreiros \cite{Vazquez-Barreiros2015} et al., who describe a local search based approach to repair a process model to include duplicate activities, starting from an event log and a process model without duplicate activities. Existing work on mining models with duplicate activities all base their duplicate activities on how well the event log fits the process model, and do not try to find any semantic difference between the multiple versions of the activities in the form of data attribute differences.\looseness=-1

The work that is closest to our work is the work by Lu et al. \cite{Lu2016}, who describe an approach to pre-process an event log by refining event labels with the goal of discovering a process model with duplicate activities. The method proposed by Lu et al., however, does not base the relabelings on data attributes of those events but instead bases them solely on the control flow context, leaving uncertainty whether two events relabeled differently are actually semantically different.

Another area of related work is data-aware process mining, where the aim is to discover rules with regard to data attributes of events that decide decision points in the process. De Leoni and van der Aalst \cite{DeLeoni2013} proposed a method that discovers data guards for decision points in the process based on alignments and decision tree learning. This approach relies on the discovery of a behaviorally well-fitting process model from the original event log. When only overgeneralizing process models (i.e. allowing for too much behavior) can be discovered from an event log, the correct decision points might not be present in the discovered process model at all, resulting in this approach not being able to discover the data dependencies that are in the event log. Our label refinements use data attributes prior to process discovery to enable discover more behaviorally constrained process models by bringing parts of the event attribute space to the event label. \looseness=-1

\section{Conclusion \& Future Work}
\label{sec:conclusion}
We have proposed a framework based on techniques from the field of circular statistics to refine event labels automatically based on their timestamp attribute. We have shown through a proof of concept on a real life event log that this framework can be used to discover label refinements that allow for discovery of more insightful and behaviorally more specific process models.
An interesting area of future work is to explore the use of other types of event data attributes to refine labels, e.g. power values of sensors. A next research step would be to explore label refinements based on multiple data attributes combined. This would bring challenge of clustering on partially circular and partially euclidean data spaces.\looseness=-1

\bibliographystyle{acm}
\bibliography{bibliography}
\end{document}